\documentstyle[preprint,aps]{revtex}
\tightenlines
\begin{document}
\draft

\title{Coherent state formulation of pion radiation from
nucleon antinucleon annihilation}

\author{R.D.~Amado$^{a,b}$, F.~Cannata$^{a,c}$,
J.-P.~Dedonder$^{a,d}$,  M.P.~Locher$^{a}$ and
{}~Bin ~Shao$^{b}$}
\address{$^a$ Paul Scherrer Institute, CH-5232 Villigen-PSI, Switzerland, \\
$^b$ Department of Physics,
University of Pennsylvania, Philadelphia,
PA 19104, USA  \footnote{Permanent address of RDA} \\
$^c$ Dipartimento di Fisica and INFN,
I-40126 Bologna, Italy \footnote{Permanent address of FC} \\
$^d$ Laboratoire de Physique
Nucl\'eaire, Universit\'e Paris 7,
2 Place Jussieu, F-75251 Paris Cedex 05 and
Division de  Physique Th\'eorique,
IPN, F-91406 Orsay, France  \footnote{Permanent addresses of JPD}}

\date{\today}

\maketitle

\begin{abstract}
We assume that nucleon antinucleon annihilation is
a fast process leading to a classical coherent pion
pulse.  We develop the quantum description of such
pion waves based on the method of coherent states.
We study the consequences of such a description for
averages
of charge types and moments of distributions
of pion momenta
with iso-spin
and four-momentum conservation taken
into account.
We briefly discuss the applicability
of our method to annihilation
at rest, where we find
agreement with experiment,
and suggest other
avenues for its use.
\end{abstract}
\newpage
\section{Introduction}

Low energy nucleon-antinucleon annihilation is a fertile
area for studying hadron dynamics and particularly
pionization.  Experiments give information on pion
numbers, spectra, and correlations and seem to show
little dependence on initial energy from annihilation at
rest to kinetic energies of a few hundred MeV.
Results do depend on the iso-spin and
spin quantum numbers of the annihilating pair.
Understanding
these results is a serious challenge to theory.

In this paper we develop a description of nucleon-antinucleon
annihilation into pions based on the creation of a
coherent classical pion wave in the annihilation
process and the quantization of this wave using the
method of coherent states. About 40\% of annihilations
go through meson resonances, mostly
rho and omega mesons, and then to pions \cite{Amsler}.  These
channels do not lead to prompt pions and are not
part of our treatment.
Our first studies show
our picture accounts for the major
phenomenological features of the pions seen in
low energy annihilation. Our picture divides the
process into two steps.  The first is the dynamics
of the nucleon and antinucleon.  We have little
to say about this except to assume that once
these two ``touch" and annihilation begins
it proceeds very rapidly.  This rapid annihilation
leads to a pion pulse or pion wave that
forms the basis of our coherent state.
It is the description and consequences of that state
and the way in which its features reflect the spatial and
temporal evolution of the annihilation region as well as its
quantum numbers that is our principal concern \cite{prl}.

Our description of annihilation as occurring by the
rapid radiation of pions, draws its inspiration
from recent studies of annihilation in the Skyrme
model. Sommermann {\it et al} \cite{Sommermann} have
studied numerically the collision of two
Skyrmions, one of baryon number one and the other
of baryon number minus one. They find that once
there is significant overlap of the two, they
disappear into pion radiation at nearly the causal
limit.  Shao {\it et al} \cite{Shao} studied this rapid
decay in a schematic model.  They considered
an initial well localized ``blob"
of Skyrmionic matter with baryon number zero and
total energy about twice the single Skyrmion energy,
and studied its time evolution.
They found that it decays into a pion wave
nearly at the causal limit. That pion wave
is a propagating, but well localized pulse.
This picture of the annihilation process
in terms of a rapid coherent radiation pulse is as
far from the thermodynamic one
as a broadcasting
radio antenna is from a black body.

The Skyrmionic description is classical, and is
meant as purely suggestive. Does a classical starting
point make sense for annihilation? Let us assume that the
rapid conversion to pion radiation seen in the Skyrme
treatments is a feature of the real annihilation.
Then the pions are
radiated from this ``bright flash" as a coherent
pulse, or coherent pion quantum wave. Since the mean
number of pions radiated is not small, and the total energy
to be radiated is large compared with the pion mass, $\mu$, we can
envisage this wave to be, in first approximation, a
{\bf classical coherent wave}. The remaining problem is
then to quantize this wave. This we do with the well known method of
coherent states. Our picture, then, has two central assumptions.
First we assume that rapid annihilation leads to
a classical pion wave pulse.  This is a consequence of the
rapid time scale of the annihilation and the large
energy (compared to the pion mass) released. Second we
assume that this classical wave can be quantized using
coherent states.  This is basically the first term
in a description with expansion parameter $1/\hat{N}$,
where $\hat{N}$ is the mean number of pions.
The error in our second assumption can therefore
be estimated.

It should be noted that our starting point arises in
classical QCD, which is the non-perturbative domain
of the theory.  This seems to us the correct starting
point for the study of annihilation. It may be
that other strongly interacting systems are also
best approached from this direction.  It has recently
been
shown that that is true for the nuclear
force \cite{WA}. In such treatments it is the
reintroduction of quantum aspects that is the
problem. For radiation fields,
coherent states are constructed to solve that problem.

In the next Section, we develop the coherent state
formalism for the pions.  We begin with a description
of classical pion waves from a source,
introduce coherent states, discuss the constraints
of iso-spin and of energy and momentum conservation and
give some statistical features of
the coherent state. In Section 3 we briefly
investigate the consequences of our description for
the annihilation data.  We only touch on what could be done,
but we do see that our approach generally agrees with experiment.
Finally Section 4 discusses some directions
for further work and presents a brief summary.

\section{Coherent State Formalism for Pions}

\subsection{Classical Pion Waves}

We imagine nucleon antinucleon annihilation as occurring very
rapidly from a small region of high energy density.
In first approximation, we take this region to radiate a
classical coherent pion wave.  If we further neglect $\pi-\pi$
interactions and the back reaction of the radiated pions on the
source,
we can write the
classical wave equation for the pion wave $\Phi({\bf r},
t)$, as
\begin{equation}
  (\nabla ^2 - \frac{\partial^{2}}{\partial t^2} - \mu ^2) \Phi({\bf r},t)
=  S({\bf r},t) .
\end{equation}
Here $S({\bf r},t)$ is the source of the pion field, $\Phi$. This
source will be assumed to be sharply localized in space and time.
For the moment we neglect the isospin degree of freedom and will return
to it later. We also leave out the factor of $-4 \pi$ in front of the
source that is familiar from electrodynamics, since it adds nothing here.
Eqn (1) gives the form of the pion wave as a classical wave radiated by
the source $S$. The details of the annihilation mechanism
are contained in this source function.

The equation for $\Phi$ is easily solved in terms of $S$ by taking
Fourier transforms.  We define the Fourier transform of $\Phi$ by
\begin{equation}
   \Phi({\bf r},t) = \int \frac{d^3k d\omega}{(2 \pi)^2} e^{i {\bf k}
\cdot {\bf r}} e^{-i \omega t} \phi({\bf k},\omega) ,
\end{equation}
and a corresponding function, $s({\bf k}, \omega)$ as the Fourier
transform of $S({\bf r},t)$. The pion field is then given by
\begin{equation}
   \Phi({\bf r},t) = \int \frac{d^3k d\omega}{(2 \pi)^2} e^{i {\bf k}
\cdot {\bf r}} e^{-i \omega t} \frac{s({\bf k},\omega)}{\omega ^2
-k^2-\mu ^2} .
\end{equation}
The pion radiation field comes entirely from the poles of the
denominator in the integral.  We must specify how
those poles are to be treated in order to get causal
radiation.  Doing so and using the pole contributions only to
evaluate the $\omega$
integral in (3), we obtain the approximate expression
for the pion field, for $t>0$,
\begin{equation}
   \Phi({\bf r},t) = -i\int \frac{d^3k}{(4 \pi \omega_k)}
 s({\bf k},\omega_k)
 e^{i {\bf k} \cdot {\bf r}  -i \omega_k t} .
\end{equation}
where $\omega_k =\sqrt{k^2 + \mu^2}$.
This form for the pion field is exact in the radiation region,
but neglects non-radiating parts of the field coming from the
$\omega$ dependence of $s$.  In the radiation zone there is no
source and $\Phi$ given in (4) satisfies
the source free wave equation.
It is this form for $\Phi$ that we will take to represent the
radiated pions of nucleon antinucleon annihilation.

\subsection{Coherent States}

It is not the classical pion wave that is detected in annihilation
experiments but rather the quanta of that field, the pions.
We must therefore extract the quantum content of this classical
wave.  This is exactly what the coherent state method achieves.
Pioneered by Glauber \cite{glauber} for optics in situations where
the photon granularity is significant but not overwhelming, the
coherent state formalism allows one to construct a quantum state
that corresponds, in the large number of field quanta limit, to
the classical wave field. Coherent states for pions were studied by
Horn and Silver \cite{HornSilver}, and many of the arguments given
here were considered by them.  They did not, however, study annihilation.

Because the pion field satisfies the wave equation without sources,
it can be decomposed into a linear superposition of free waves
with wave number ${\bf k}$ and corresponding energy $\pm \omega_k$.
That is the content of Eqn (4).  We can bring in quantum mechanics
for this wave by introducing the standard creation and annihilation
operators for the modes ${\bf k}$, $a_{{\bf k}}$ and $a^{\dagger}_
{{\bf k}}$. These satisfy the usual harmonic oscillator commutation
relations,
\begin{equation}
[a_{{\bf k}}, a^{\dagger}_{{\bf k'}}] = \delta({\bf k}-{\bf k'}),
\end{equation}
with all other commutators zero.
The quantum field operator corresponding to the pion field is then
\begin{equation}
\Phi_{qm}({\bf r},t)=\int \frac{d^3k}{(2\pi)^{3/2}}
( a_{{\bf k}} e^{i {\bf k} \cdot {\bf r}} e^{-i \omega_k t} +
  a^{\dagger}_{{\bf k}} e^{-i {\bf k} \cdot {\bf r}}
 e^{i \omega_k t})
\end{equation}
The coherent state associated with a given classical state,
$\Phi({\bf r},t)$, is
the quantum state that is an eigenstate of
the positive frequency part of   $\Phi_{qm}$, $\Phi_{qm}^+$,
with the corresponding positive energy part of the
classical wave as its eigenvalue.
Recall that the positive frequency part of the
state has the $e^{-i \omega_k t}$ dependence.
In the quantum operator for the field, this is the part that
goes with the annihilation operator.  Since the modes for each
${\bf k}$ are independent, finding the eigenstate is equivalent
to finding the eigenstate of one single annihilation operator.
The full state can then be built up by superposing the independent
modes.

The normalized eigenstate, $|\lambda \rangle $, with eigenvalue,
$\lambda$,
of a single mode annihilation operator, $a$,
( $ a |\lambda\rangle  = \lambda |\lambda\rangle $), is given by
\begin{equation}
|\lambda\rangle  = e^{-\frac{\lambda \lambda^{\ast}}{2}}
e^{\lambda a^{\dagger}}|0\rangle
\end{equation}
where $|0\rangle $ is the vacuum state.  This equation for
the eigenstate of $a$ is the key  to the entire coherent state
formalism.  \footnote{ We are talking here of field coherent
states.  These are the coherent states appropriate to
classical fields. They are different from
coherent states expressed as polynomials in the
creation operators and
used in descriptions of material
objects such as atoms, molecules, deformed nuclei and
Skyrmions. (For an introduction to these states see \cite{Gin})}

Consider now the normalized quantum state $|f\rangle $ defined by
\begin{equation}
|f\rangle  = \exp(-\frac{1}{2}\int d^3k f^{\ast}({\bf k})
f({\bf k}) + \int d^3k f({\bf k}) a^{\dagger}_
{{\bf k}} ) |0\rangle  .
\end{equation}
This is just a coherent state in which each mode ${\bf k}$ carries
weight $f({\bf k}) d^3k$.  It is clear, for the positive frequency
part of the field, that
\begin{equation}
\Phi_{qm}^{+}({\bf r},t) |f\rangle  = \varphi({\bf r},t) |f\rangle ,
\end{equation}
with the eigenfunction $\varphi$ given by
\begin{equation}
\varphi({\bf r},t) = \int \frac{d^3k}{(2 \pi)^{3/2}} e^{i {\bf k}
\cdot {\bf r}} e^{-i \omega_k t} f({\bf k}).
\end{equation}
This $\varphi$ satisfies the source free wave equation
for any function $f$. We then take $|f\rangle $ to be the normalized quantum
state corresponding to the classical wave $\varphi$.  Note that
$|f\rangle $  does
not contain a fixed number of quanta, but is rather a coherent
superposition of states of different numbers of quanta.

The meaning of a coherent state arising from a classical source
can be clarified by relating it to the standard S-matrix description
of a field generated by a classical external source. The classical
source, $S$, coupled to the field, $\Phi$, creates from the vacuum the
quantum state
\cite{Itzykson}
\begin{equation}
\exp(-i \int d^3r dt \Phi_{qm}({\bf r},t) S({\bf r},t)) |0\rangle .
\end{equation}
Upon four dimensional Fourier
transformation and integration over $d k_o $, this reduces,
up to a normalization, to Eqn (8). The exponent in (11) is
just the S-matrix  in the interaction representation, and
makes clear both the connection of the S-matrix and coherent state
approaches and why the resulting state has an indefinite number
of quanta.

In the electromagnetic
case it seems natural to define states with indefinite numbers
of photons, but even there if the wave has a fixed total energy
only a fixed number of photons above some frequency can be
detected. With pions the description in terms of a coherent mixture
of states with different numbers of pions seems less comfortable,
particularly in view of the finite pion mass.  No one
would suggest representing a pion wave with total energy
near the pion rest mass as a classical object.
In annihilation, however, the total energy released is about
13 times the pion rest mass, and the approximation of
representing it by a classical coherent wave seems
at least plausible.  The purpose of this paper is to show that
that description is not only plausible
but useful and gives dynamical insight into
the annihilation process and may have application in a number
of other processes in which many pions are created.

To make the pion coherent state correspond to the radiating
solution $\Phi({\bf r},t)$ of Eqn (4), that is to make $\varphi$ in
(9) or (10) correspond to the positive frequency part of $\Phi$,
we must take in (8)
\begin{equation}
 f({\bf k}) = -i \frac{s({\bf k},\omega_k)}{2 \omega_k} \sqrt {2 \pi} .
\end{equation}
With this choice, our quantum coherent state
represents  the classical radiating pion wave generated
by the source $S({\bf r},t)$.

\subsection{Iso-spin}
The pion is an iso-vector and comes in three charge states.
In principle each charge state can have its independent
source, corresponding
classical wave, and coherent state.
This is most easily formulated by making $\Phi({\bf r},t)$
and $S({\bf r},t)$ of Eqn (1) each iso-vectors.
Then each charge state could have a source density that
varies arbitrarily in space and time. In fact nucleon antinucleon
annihilation occurs in a state of fixed total iso-spin, 0 or 1.
The full pion wave must have the same
total iso-spin.  This condition of total iso-spin
conservation imposes a {\bf global} condition on the source and
on the coherent state. Since the different ${\bf k}$ modes
of the wave are independent, they must each combine to the
correct total iso-spin, which means that the iso-spin
source density cannot depend on ${\bf r}$ or $t$.
In the coherent state $|f\rangle $ of (8) we expect in general
that $f({\bf k})$ is an iso-vector dotted into the now
iso-vector creation operator.  But if we are to construct
states of fixed iso-spin for any ${\bf k}$ configuration,
the iso-vector dependence of
$f$ must be independent of ${\bf k}$.  In that case
$f$ is a constant vector in iso-spin space. The direction
of that iso-vector depends on the source.

For the moment
let us consider a coherent state specified by a fixed
unit vector in iso-spin space $\hat{T}$ and an iso-scalar
function $f({\bf k})$.  Call that state $|f,\hat{T}\rangle $.
In place of (8) we write
\begin{equation}
|f,\hat{T}\rangle
 = \exp(-\frac{1}{2}\int d^3k f({\bf k})
f^{\ast}({\bf k}) + \int d^3k
\hat{T} \cdot {\bf a}^{\dagger}_
{{\bf k}} f({\bf k})) |0\rangle  .
\end{equation}
That is a
coherent state constructed from a particular
superposition of pions,
that superposition made up of pions  all pointing
in the iso-direction of $\hat{T}$.  This is not yet a state of
fixed iso-spin. In fact it is easy to see that it contains
states of all iso-spin.  A state of fixed iso-spin must
be projected from it. Call the projected state of total
iso-spin $I$ with z-component $I_z$, $|f,I,I_z\rangle $.  The
projection is given by \cite{Doug}
\begin{equation}
   |f,I,I_z\rangle  = \nu \int \frac{d\hat{T}}{\sqrt{4 \pi}}
 |f,\hat{T}\rangle  Y_{I,I_z}^{\ast}(\hat{T})
\end{equation}
where $ Y_{I,I_z}(\hat{T})$ is the usual spherical harmonic,
and $\nu$ is a normalization constant.
The state defined in (14)  is
orthogonal to states of different total iso-spin or
z-component projection.  To see this and to calculate
the normalization
let us study the inner product of two projected
states.  This is given by
\begin{equation}
\langle f,I,I_z|f,I',I'_z\rangle = \nu^2
 \int \frac{d\hat{T} d\hat{T'}}{4 \pi}
\langle f,\hat{T}|f,\hat{T'}\rangle  Y_{I,I_z}(\hat{T})
 Y^{\ast}_{I',I'_z}(\hat{T'})
\end{equation}
It is easy to show that
\begin{equation}
\langle f,\hat{T}|f,\hat{T'}\rangle =\exp\left(
-\frac{(\hat{T}-\hat{T'})^2}{2}
\int d^3k f({\bf k}) f^{\ast}({\bf k})\right)
\end{equation}
For a strong field, large $f$, this inner product is very
sharply peaked around $\hat{T}=\hat{T'}$.
We will show in the next section that
\begin{equation}
\int d^3k f({\bf k}) f^{\ast}({\bf k})
=\hat{N}
\end{equation}
where $\hat{N}$ is the mean number of pions emitted in the
annihilation.  We will use this notation here both to save
writing and to remind us that this integral, or $\hat{N}$,
is large.  To show the orthogonality in (16) write
\begin{equation}
e^{\hat{T} \cdot \hat{T'} \hat{N}}
=4 \pi \sum_{l,m} i^l j_l (-i \hat{N}) Y^{\ast}_{l,m}(\hat{T})
Y_{l,m}(\hat{T'}).
\end{equation}
{}From this it is easy to show that
\begin{equation}
\langle f,I,I_z|f,I',I'_z\rangle = \nu^2
e^{- \hat{N}} i^I j_I(-i \hat{N}) \delta_{I,I'}
\delta_{I_z,I'_z}
\end{equation}
from which the normalization can be read off.  In particular
for large $\hat{N}$, we can use the asymptotic form of the
Bessel function to yield for the normalization
\begin{equation}
\nu = \sqrt {2 \hat{N}}.
\end{equation}

We should note that our iso-spin projection has the
interesting property that states of even iso-spin
contain only even numbers of pions
and states of odd iso-spin only odd numbers.
To escape this restriction, as the data does, we need
to include the channels for annihilation into
meson resonances.

\subsection{Probing the Coherent State: No Iso-spin}

We would now like to examine the
properties of the coherent pion state generated by the
assumed initial source $S({\bf r},t)$. We begin with a
discussion of the case without iso-spin.
This  coherent state is given by (8)
with $f({\bf k})$ given in terms of the Fourier transform of
$S$ as in (12).
Although the state does not contain a fixed
number of pions, we can ask what is the average number,
$\hat{N}$.   This is just
the expectation value of the total pion number
operator in the coherent state.  We need
\begin{equation}
\hat{N} =
  \langle f| \int d^3k  a_{{\bf k}}^{\dagger} a_{{\bf k}} |f\rangle  .
\end{equation}
(Note that we are working
in the Heisenberg representation so that the states are
time independent while the operators
in general carry a time
dependence. However, the number operator is time independent.)
This expectation value is easily evaluated using the
special properties of the coherent states to give
(17). Thus we see that the integral that enters
to normalize the coherent state in (8), is just
$\hat{N}$. It is large $\hat{N}$
that corresponds to large field ($f$ or $S$ large)
and thus to a good approximate correspondence between
the classical and quantum descriptions.
It is also clear from this that the average single
pion momentum distribution is given by
\begin{equation}
\frac{dN({\bf k})}{d^3k} =  f^{\ast}({\bf k}) f({\bf k}).
\end{equation}
in terms of the square of the Fourier transform of the
source density.

In the same way one obtains for the total energy
released, the expectation of the Hamiltonian, or
\begin{equation}
E= \int d^3k
\omega_k  f^{\ast}({\bf k}) f({\bf k}).
\end{equation}
In a quantum process, this energy is sharp and should
be put equal to the total energy released in the
annihilation.  This fact serves as a convenient way
to normalize $f$.  The sharpness of the state can
be seen by calculating the dispersion in the number
of pions emitted.  One finds
\begin{equation}
 \sigma^2 = \langle f| N^2- (\hat{N})^2|f\rangle  = \hat{N} ,
\end{equation}
which shows that the fractional dispersion in $N$ is
$1/\sqrt{\hat{N}}$. This result also follows from the fact that
the probability of finding a state of $m$ pions is given by
a Poisson distribution, as we shall show below.

One can use the coherent state to calculate the amplitude
for finding some particular quantum configuration.
For example the amplitude for finding $m$
pions of momenta ${\bf p}_1...{\bf p}_m$ is given by
\begin{equation}
\langle {\bf p}_1,...{\bf p}_m|f\rangle =\frac{1}{\sqrt{m!}}
f({\bf p}_1)...
f({\bf p}_m) e^{-\hat{N}/2}
\end{equation}
The probability of finding $m$ pions of any momentum is
then
\begin{eqnarray}
P_m & =& \frac{1}{m!} e^{-\hat{N}} (\int |f({\bf p})|^2 d^3p)^m
\\ \nonumber
 &=& \frac{1}{m!} e^{-\hat{N}} \hat{N}^m
\end{eqnarray}
which is the Poisson distribution.
It is clear that
the coherent state does not ``know" not to emit more
pions than energy conservation permits.
We will turn to these constraints of
energy and momentum conservation
below. Our discussion here
serves to emphasize that the coherent
state approach is better at giving average
information over the ensemble of annihilations
than it is for particular quantum states.

\subsection{Probing the Coherent State: With Iso-spin}

If we include iso-spin projection as in Section 2.3,
we can also probe questions about iso-spin populations.
For example let us study the average number of pions
of a particular type in a projected coherent state
with total iso-spin $I$ and z-component $I_z$.  This
is given by
\begin{equation}
  \hat{N}_{\mu} =
  \langle I,I_z,f| \int d^3k a_{{\bf k},\mu}^{\dagger}
a_{{\bf k},\mu}
  | I,I_z,f\rangle ,
\end{equation}
where $\mu$ is the pion type ($+,-,0$) and there is no
sum over $\mu$ implied. Using the forms of the projected
states, this becomes
\begin{equation}
  \hat{N}_{\mu} =
\nu^2 \int \frac{d\hat{T} d\hat{T'}}{4 \pi} Y^{\ast}_{I,I_z}
(\hat{T})
\hat{T}_{\mu}^{\ast} \hat{T'}_{\mu} Y_{I,I_z}(\hat{T'})
exp(\hat{N}(\hat{T} \cdot \hat{T'}-1))
 \int d^3k f^{\ast}({\bf k}) f({\bf k})
\end{equation}
Using (18) and expressing the $\hat{T}$ in terms of spherical
harmonics, the integrals can be calculated in terms of standard
identities.  We continue to use (17)
to express the integral over $f$ in terms of $\hat{N}$.  We now
use (17) to define $\hat{N}$ but as we shall see it is still
the average number of pions summed over all pion types.
We then find
\begin{equation}
  \hat{N}_{\mu} =
 \hat{N} \sum_{l,m} \frac{2l+1}{2I+1}(\langle l0,10|l1,I0\rangle
\langle lm,1\mu|l1,II_z\rangle )^2
\end{equation}
in terms of a sum over Clebsch-Gordan coefficients.
In obtaining (29) we have used the asymptotic forms
of the Bessel functions
(large $\hat{N}$) so that there is no $l$
dependence in the normalization.

For nucleon-antinucleon
annihilation there are only two choices for $I$, 0 or 1.
For the case of $I=0$ (29) reduces to
\begin{equation}
  \hat{N}_{\mu} = \hat{N}/3 ,
\end{equation}
for any $\mu$.
As we expect for a state of iso-spin zero the average
number of pions is type independent, and each is
one third of $\hat{N}$, making
$\hat{N}$ the true average
number of pions summed over types, as
promised.  For $I=1$ there are three cases.  For
$I_z=0$ one finds
\begin{eqnarray}
\hat{N}_+ &=& \hat{N}/5 ~,
\nonumber \\
\hat{N}_- &=& \hat{N}/5 ~,
\nonumber \\
\hat{N}_0 &=& 3 \hat{N}/5 ~.
\end{eqnarray}
We see that again the sum of averages over types is $\hat{N}$.
Note the dominance of $\pi^0$, an important feature of the data.
For $I_z=1$ one finds
\begin{eqnarray}
\hat{N}_+ &=& 2 \hat{N}/5 ~,
\nonumber \\
\hat{N}_- &=& 2 \hat{N}/5 ~,
\nonumber \\
\hat{N}_0 &=&  \hat{N}/5 ~.
\end{eqnarray}
with the same result for $I_z=-1$.
It is no surprise that for the $I=0$ case or for
the $I=1$, $I_z=0$ case the average number of
$\pi^+$ and $\pi^-$
is the same.  It is a bit more puzzling for $I=1$,
$I_z=\pm 1$.  What happened to charge conservation?
The initial $I_z=\pm 1$ state has charge $\pm 1$. However,
$\hat{N}_{\pm}$ is a number of order $\hat{N}$.
To that order all three $I_z$ values do correspond to states
of average charge
zero.  Charge conservation is a $1/\hat{N}$
effect, and we are neglecting terms of that order.
We will return to this approximation below.

The same charge ratios can be obtained in (28) by
making the overlap of (16) proportional to
$\delta(\hat{T}-\hat{T'})$.  We have already commented that this
overlap is strongly peaked around $\hat{T}=\hat{T'}$ for
large $\hat{N}$.  This delta function approximation is
correct in the large $\hat{N}$ limit and it makes
calculations of the average pion number
by type and of higher moments of the distribution
very simple.

We illustrate this delta function method
in the calculation of
isospin number correlations.
For example the joint average
of $\mu$ type and $\nu$ type pions in a state of total
iso-spin $I$ with z-component $I_z$ is
\begin{equation}
\hat{N}^{2}_{\mu,\nu}=
 \langle I,I_z,f|\int d^3 k a^{\dagger}_{{\bf k},\mu} a_{{\bf k},\mu}
 \int d^3 p a^{\dagger}_{{\bf p},\nu} a_{{\bf p},\nu}
 |I,I_z,f\rangle .
 \end{equation}
 In the large $\hat{N}$ limit where the overlap may be
 replaced by a delta function, we obtain,
 \begin{equation}
 \hat{N}^{2}_{\mu,\nu}= \hat{N}^2
 \int d\hat{T} Y_{I,I_z}^{\ast}(\hat{T})
 \hat{T}^{\ast}_{\mu} \hat{T}_{\mu}
 \hat{T}^{\ast}_{\nu} \hat{T}_{\nu}
  Y_{I,I_z}(\hat{T}).
  \end{equation}
In this expression the components of $\hat{T}$
are, in the usual spherical coordinates,
\begin{eqnarray}
\hat{T}_0 &=& \cos \theta
\nonumber \\
\hat{T}_+ &=& -\frac{1}{\sqrt{2}} \sin \theta e^{i \phi}
\nonumber \\
\hat{T}_- &=& \frac{1}{\sqrt{2}} \sin \theta e^{-i \phi} .
\end{eqnarray}
Higher order forms are obtained from
the generalization of (33) and (34).  They come from
higher powers of the number operator for the $\rho$th
type in the expectation value with the coherent state
replacement
\begin{equation}
 \int d^3 k a^{\dagger}_{{\bf k},\rho} a_{{\bf k},\rho}
 = \hat{N} \hat{T}^{\ast}_{\rho} \hat{T}_{\rho}
 \end{equation}
 The integrals that appear in (34) are elementary, and
 hence these higher moments are easily calculated.

Greiner, Gong and Muller \cite{Greiner}
have recently introduced polynomial coherent states in a discussion of
pion condensates in heavy ion collisions.  They consider only
$I=0$ states and calculate the iso-spin pair correlations
defined as
\begin{equation}
C_{\mu,\nu}=\frac{\hat{N}^{2}_{\mu,\nu}}{\hat{N}_{\mu}
\hat{N}_{\nu}} -1.
\end{equation}
We obtain the same values as they do for
$C_{\mu,\nu}$ for large $\hat{N}$, but our method is
simpler.  We can also use the method to
calculate the pair correlations in other states.  For example
we find for $I=1$, $I_z=0$, $C_{0,0}=4/21$ and for $I=1$,
$I_z=1$, $C_{0,0}= 8/7$.  Other moments and other amplitudes
are also easily calculated and agree with those of Greiner,
Gong and Muller where calculated by them.

Thus we see that the coherent state approach makes
definite predictions about average charge ratios
in annihilations and can also be used to calculate
higher moments of those distributions.  It also
predicts that the single pion momentum distribution
should be independent of pion type.

\subsection{Energy and Momentum Conservation}

So far the coherent state we have defined does not have
definite total energy or momentum. For example it contains
contributions from states with only one pion or with
arbitrary numbers of pions,
neither of which is allowed by conservation of total energy
and momentum.  This fault is not serious for large $\hat{N}$,
but is significant for the intermediate $\hat{N}$ of
annihilation.  Energy and momentum conservation can be imposed
on the coherent state, as was shown by Horn and Silver
\cite{HornSilver}, at the expense of some further complication
of the formalism.

Let us begin by imposing energy and momentum conservation on the
coherent state without iso-spin.
The method for imposing energy and
momentum on the coherent state is formally the same as the projection
method for imposing fixed iso-spin.
The projections commute, but taking them
together complicates the physics.
For iso-spin one first constructs a state with all the pions
pointing
in the same direction in iso-spin and then
averages over that direction  with the appropriate $SU(2)$
eigenfunction weight.  To construct a  state of fixed energy
and momentum (or four-momentum) one first constructs a state
of all the pions at a fixed place in space and time and averages
over those places and times with the appropriate eigenfunctions of
definite energy and momentum.  To do this define the
operator $F({\bf r},t)$ by
\begin{equation}
F({\bf r},t) = \int d^3p f({\bf p}) a_{{\bf p}}^{\dagger}
\exp(i {\bf p} \cdot {\bf r}-i \omega_p t).
\end{equation}
Then the (un-normalized) coherent state at ${\bf r}$ and $t$ is
\begin{equation}
|f,{\bf r},t\rangle  =e^{F({\bf r},t)}|0\rangle .
\end{equation}
It is more concise as well as more physical to use four
dimensional notation.  We call $x$ the four-vector of position
with space part ${\bf r}$ and time part $t$.
Thus $|f,{\bf r},t\rangle $ becomes $|f,x\rangle $.
Then the state of fixed
total four-momentum, $K$ is given by
\begin{equation}
|f,K\rangle  = \int \frac{d^4x}{(2 \pi)^4} e^{-i K \cdot x} |f,x\rangle .
\end{equation}
To see how this works and to calculate the normalization,
consider the overlap of two states of different four-momentum. One
easily finds
\begin{equation}
\langle f,K'|f,K\rangle  =\int\frac{d^4xd^4x'}{(2 \pi)^8}
\exp(-i K \cdot x +i K' \cdot x')
e^{\rho(x- x')}
\end{equation}
where
\begin{equation}
\rho (x)=\int d^3p |f({\bf p})|^2 e^{-i p \cdot x}
\end{equation}
and where the fourth component of $p$ under the integral is
$\omega_p =\sqrt{p^2+\mu^2}$.  A simple change of variable
reduces the overlap to
\begin{equation}
\langle f,K'|f,K\rangle  =\delta ^4(K-K') \int \frac{d^4x}{(2 \pi)^4} e^{i K
\cdot x}
e^{\rho(x)}
\end{equation}
This clearly shows that states of different four-momentum are
orthogonal, but the remaining integral is singular due to the
undamped large $x$ behavior of the integrand.
Horn and Silver \cite{HornSilver} also pointed out this problem.
It arises from the first term in the expansion of the exponential.
This term, a ``one", corresponds to the no pion contribution to
the annihilation process.  Since there is in fact no such
contribution that conserves energy and momentum, we can subtract
it out.  Similarly there is no one pion state that can contribute
to annihilation and conserve four-momentum, thus we may also
subtract out the one meson state and define a coherent state
of total four-momentum $K$
that starts with two pions.  This is given in terms of
\begin{equation}
|f,x,2\rangle  = (e^{F(x)} -F(x)-1) |0\rangle
\end{equation}
by
\begin{equation}
|f,K,2\rangle  = \int \frac{d^4x}{(2 \pi)^4} e^{-i K \cdot x} |f,x,2\rangle .
\end{equation}
We then find
\begin{equation}
\langle f,K',2|f,K,2\rangle  =\delta ^4(K-K') \int \frac{d^4x}{(2 \pi)^4} e^{i
K \cdot x}
(e^{\rho(x)}-\rho(x)-1).
\end{equation}
The integral now is convergent and we have lost nothing by the subtractions
since we have only removed states that cannot contribute physically.
However the integral in (46) is still quite delicate and difficult to
evaluate numerically because it contains delta functions coming from
the $d^4x$ integral.  We can get around this problem and shed more physical
light on the expression by expanding the exponential and interchanging
the order of integration.  Let us call the integral in (46) ${\cal I}(K)$.
We can write
\begin{eqnarray}
{\cal I}(K)& =& \int \frac{d^4x}{(2 \pi)^4} e^{i K \cdot x}
\sum_{m=2}\frac{\rho^m(x)}{m!},
\nonumber \\
&=& \sum_{m=2}{ \cal I}_m(K)/m! .
\end{eqnarray}
Using the definition of $\rho$ and interchanging integrations one finds
\begin{equation}
{\cal I}_m(K)=\int \delta^4(K-p_1-p_2...-p_m) \prod_{i=1}^m d^3p_i
|f({\bf p}_i)|^2,
\end{equation}
where as before we take the fourth component of $p_i$ to be
$\sqrt{p_i^2+\mu^2}$. Integrals of this form are easily done
by Monte Carlo methods
following a program given by Barger and Phillips.
\cite{B&P}

The individual terms in the $m$ sum
in (47) represent the
relative contributions to
annihilation into $m$ pions. For fixed total energy the sum
must terminate.  For example, nucleon-antinucleon annihilation at rest
cannot go into more than 13 pions, and in that case
${\cal I}_m$ must be zero for $m > 13$.  In fact ${\cal I}_m/
m!{\cal I}$ is just
the fraction of annihilations that  go into $m$ pions.
Thus the mean pion number is given by
\begin{equation}
\hat {N} = \sum _{m=2} \frac{m {\cal I}_m(K)}{m! {\cal I}(K) }
\end{equation}
This is equal to the expression
that one arrives at by taking the
expectation value of the number operator in the
state $|f,K,2\rangle $ and normalizing.
This expression can also be written
\begin{equation}
\hat{N} = \int \frac{d^4x}{(2 \pi)^4} e^{iK \cdot x} \rho(x)
(e^{\rho(x)} -1) / \int \frac{d^4x}{(2 \pi)^4} e^{iK \cdot x}
(e^{\rho(x)}-\rho(x)-1)
\end{equation}
Note that $\rho(0)= \hat{N}$ in terms of the old expression for
$\hat{N}$, (17).  For $\hat{N}$ large, $\rho(x)$ is large, and
the integrands in (50) are dominated by the factors of $e^{\rho}$
and $e^{-iK \cdot x}$. The
expression (50) can then be thought of in terms of a ratio of
improper integrals
\begin{equation}
\hat{N} \simeq  \int \frac{d^4x}{(2 \pi)^4} e^{iK \cdot x} \rho(x)
e^{\rho(x)}  / \int \frac{d^4x}{(2 \pi)^4} e^{iK \cdot x}
e^{\rho(x)}
\end{equation}
These integrals can be approximated by the method of stationary
phase.
For annihilation at rest, the total energy released, $K_0=E$,
is large, and then
the stationary phase point comes at $x \simeq 0$.
Thus the $\rho(x)$ in the numerator of (51), but not
in the exponent, can be evaluated
at $x=0$ to give that the new $\hat{N}$ is approximately equal to
the old one.  Corrections are of order $1/\hat{N}$.  We have verified
this in a number of numerical examples.

We see that four-momentum conservation
can indeed be implemented for the coherent state, but for average
quantities, and for $\hat{N}$ large, that constraint is not important.
As we will see below, it is important for the probability
distribution  and for higher moments of the distribution.
It is clear that one can combine the method outlined above
for projecting onto states of good four-momentum with the
method of the previous section for projecting onto states of
good iso-spin. We now turn to a discussion of how to do that.

\subsection{Probing the Coherent State: With Iso-spin and
Four-momentum}

Combining the methods of previous sections, we can construct a
state of fixed total four-momentum $K$ and fixed iso-spin
$I$ with z-component $I_z$. We first generalize $F$ of (38) to give
\begin{equation}
F(x,{\bf T}) = \int d^3p f({\bf p})e^{-ip \cdot x} a^{\dagger}_{{\bf p},
\mu} T_{\mu}
\end{equation}
Then the appropriate generalization of (44) to include
iso-spin is
\begin{equation}
|f,x,{\bf T},2\rangle =(e^{F(x,{\bf T})} -F(x,{\bf T})-1)|0\rangle
\end{equation}
giving the state (not normalized)
\begin{equation}
|f,K,I,I_z,2\rangle  = \int \frac{d^4x}{(2 \pi)^4} \frac{d {\bf T}}{\sqrt{4
\pi}}
e^{i K \cdot x} |f,x,{\bf T},2\rangle  Y^{\ast}_{I,I_z} ({\bf T}).
\end{equation}
The inner product of two such states belonging to different
four-momentum and iso-spin gives a four dimensional delta function
on four-momentum and Kronecker deltas on iso-spin and its z-component
times the following normalization integral
\begin{equation}
{\cal I}=\int \frac{d^4 x}{(2\pi)^4} \frac{d\hat{T} d\hat{T}^{'}}{4\pi}
e^{iK\cdot x}Y^{*}_{II_z}(\hat{T})Y_{II_z}(\hat{T}^{'})(e^{\rho(x)
\hat{T}\cdot\hat{T}^{'}}
-\rho(x)\hat{T}\cdot\hat{T}^{'}-1), \label{eq:normal}
\end{equation}
This integral can be calculated by the same expansion methods
used in (46) and (47).  In terms of the ${\cal I}_m(K)$ defined in (47)
the normalization integral of (\ref{eq:normal}) is given by
\begin{equation}
{\cal I}=\sum_{m=2}\frac{{\cal I}_{m}(K)}{m!}F(m,I), \label{eq:Fmsum}
\end{equation}
where
\begin{equation}
F(m,I)=\int  \frac{d\hat{T} d\hat{T}^{'}}
{4\pi} Y^{*}_{II_z}(\hat{T})Y_{II_z}(\hat{T}^{'})
(\hat{T}\cdot\hat{T}^{'})^{m}
\end{equation}
This integral can be done using the identity in (18) by
differentiating (18) with respect to $\hat{N}$ $m$ times and then
setting $\hat{N}$ to zero. One finds
\begin{equation}
F(m,I)=\left\{   \begin{array}{ll}
0 & I >  m \mbox{ and } I-m \mbox{ is odd} \\
\frac{ m! }
{ (m-I)!! (I+m+1)!! } & I\le m \mbox{ and }
I-m \mbox{ is even}.
\end{array}   \right.
\end{equation}
It should be recalled that sums over $m$
as in (\ref{eq:Fmsum}) have finite upper limits because of the constraint
of energy conservation.

The expected number of pions of type $\mu$ corresponding to
(27) is now
\begin{equation}
\hat{N}_{\mu}=\frac{1}{\cal{I}}\int \frac{d^4 x}{(2\pi)^4}
\frac{d\hat{T} d\hat{T}^{'}}{4\pi}
e^{iK\cdot x}Y^{*}_{II_z}(\hat{T})Y_{II_z}(\hat{T}^{'})
\hat{T}_{\mu}\hat{T}^{'}{}^{*}_{\mu}\rho(x)
(e^{\rho(x)\hat{T}\cdot\hat{T}^{'}}-1),  \label{eq:number}
\end{equation}
The joint average of $\mu$ type and $\nu$ type pions as
in (33) is given by
\begin{equation}
\hat{N}^{2}_{\mu\nu}=\frac{1}{\cal{I}}
\int \frac{d^4 x}{(2\pi)^4} \frac{d\hat{T} d\hat{T}^{'}}{4\pi}
e^{iK\cdot x}Y^{*}_{II_z}(\hat{T})Y_{II_z}(\hat{T}^{'})
\hat{T}_{\mu}\hat{T}^{'}{}^{*}_{\mu}
\hat{T}_{\nu}\hat{T}^{'}{}^{*}_{\nu}
\rho^{2}(x) e^{\rho(x)\hat{T}\cdot\hat{T}^{'}}  \label{eq:corr}
\end{equation}
In obtaining (\ref{eq:corr}), we have neglected a term from the commutator
that is down by $1/\hat{N}$. In these expressions, $\rho$ is still
given as in (42).
These expectation values can be evaluated, as before, by expanding.
Once again they are expressed in terms of the ${\cal I}_m$.  We have
\begin{equation}
\hat{N}_{\mu}=\frac{1}{\cal{I}}\sum_{m=2}\frac{{\cal I}_{m}(K)}{(m-1)!}
G_{\mu}(m-1,I,I_z),
\end{equation}
and
\begin{equation}
\hat{N}^{2}_{\mu\nu}=\frac{1}{\cal{I}}
\sum_{m=2}\frac{{\cal I}_{m}(K)}{(m-2)!}H_{\mu\nu}(m-2,I,I_z),
\end{equation}
where
\begin{eqnarray}
&& G_{\mu}(m,I,I_z)=
\int  \frac{d\hat{T} d\hat{T}^{'}}
{4\pi} Y^{*}_{II_z}(\hat{T})Y_{II_z}(\hat{T}^{'})
\hat{T}_{\mu}\hat{T}^{'}{}^{*}_{\mu} (\hat{T}\cdot\hat{T}^{'})^{m} \\ &&
=\sum_{ln}F(m,l) \frac{2l+1}{2I+1}(\langle l0,10|I0\rangle
\langle ln,1\mu|II_z\rangle )^2 \nonumber
\end{eqnarray}
and
\begin{eqnarray}
&& H_{\mu\nu}(m,I,I_z)=
\int  \frac{d\hat{T} d\hat{T}^{'}}{4\pi}
Y^{*}_{II_z}(\hat{T})Y_{II_z}(\hat{T}^{'})
\hat{T}_{\mu}\hat{T}_{\nu}\hat{T}^{'}{}^{*}_{\mu}\hat{T}^{'}
{}^{*}_{\nu}(\hat{T}\cdot\hat{T}^{'})^{m}
\\ &&
=F(m,I)(\langle 10,10|00\rangle \langle 1\mu,1\nu|00\rangle )^2
+ 2F(m,I)\langle 10,10|00\rangle \langle 1\mu,1\nu|00\rangle
\nonumber \\ &&
\langle 10,10|20\rangle \langle I0,20|I0\rangle
\langle 1\mu,1\nu|2\mu+\nu\rangle \langle II_z,2\mu+\nu|II_z\rangle
\nonumber \\ &&
+\sum_{ln}F(m,l)\frac{2l+1}{2I+1}(\langle 10,10|20\rangle \langle
l0,20|I0\rangle
\langle 1\mu,1\nu|2\mu+\nu\rangle \langle ln,2\mu+\nu|II_z\rangle )^2 \nonumber
\end{eqnarray}

Let us examine these forms in the cases of physical interest.
For $I=0$, it is easy to show that
\begin{equation}
G_0(m,0,0)=G_1(m,0,0)=G_{-1}(m,0,0)=F(m,1)/3
\end{equation}
Then from (56) and (61) we get
\begin{equation}
\hat{N}_+=\hat{N}_-=\hat{N}_0 = \hat{N}
\end{equation}
as we expect.  For $I=1$, $I_z=0$, we find
\begin{eqnarray}
G_0(m,1,0)&=& F(m,0)/3 +4F(m,2)/15  \nonumber \\
G_1(m,1,0)&=& F(m,2)/5  \nonumber \\
G_{-1}(m,1,0)&=& F(m,2)/5.
\end{eqnarray}
These do not seem to give the same charge ratios
we obtained before in this state, but those were
calculated in
the large $\hat{N}$ limit.  In that limit the
dominant terms in the sums of (56) and (61) come from
terms with $m \sim \hat{N}$.  Thus the dominant terms
in $F(m,l)$ that matter have $m >> l$ (recall that $l$ is
of order $I$ which is in turn of order $1$). In this limit it
is easy to show that $F(m,l)/F(m,l+2) \sim 1$ with corrections
of order $1/m$.  In that limit (67) yields the results of (31).
Similarly for $I=1$, $I_z=1$, we find
\begin{eqnarray}
G_0(m,1,1)&=& F(m,2)/5  \nonumber  \\
G_1(m,1,1)&=& F(m,0)/3+F(m,2)/15   \nonumber \\
G_{-1}(m,1,1)&=& 2F(m,2)/5.
\end{eqnarray}
which yields (32) in the large $\hat{N}$ limit,
and shows how charge conservation enters for finite
$\hat{N}$. One can calculate the two pion
averages in the same way.  The factors that enter
become for $I=1$, $I_z=0$
\begin{equation}
H_{00}=9F(m,1)/25 +12F(m,3)/175
\end{equation}
and for $I=1$, $I_z=1$
\begin{equation}
H_{00}=F(m,1)/25 +8F(m,3)/175
\end{equation}
which yield our previous results for the
iso-spin pair correlations $C_{\mu,\nu}$ defined
in Sect. 2.5 in the large $\hat{N}$ limit.

The evaluation of the averages now depend on the functional form of
$f({\bf k})$. We turn to it in the next section.

\section{ Phenomenology}

As a first orientation into the phenomenological content of
the coherent pion wave description of annihilation, let us
study the relation between the average pion number
$\hat{N}$ of (17) and the energy released, (23), for
a very simple model of $f$.
We do not take $f$ directly from the experimental pion spectrum
since the pions from the meson resonance channels confuse
the extraction.  Rather
suppose we assume that the pion field source turns on
at $t=0$ and then decays exponentially in time, and
that it has a spherically symmetric
Yukawa shape. (Spherical symmetry is appropriate for
annihilation at rest.) That is we take
for $S({\bf r},t)$ of (1),
\begin{eqnarray}
 S({\bf r},t) &=& 0 ,\hspace{.2in} t < 0  \\ \nonumber
   &=& S_0 \frac{e^{-\alpha r}}{r} t e^{-\gamma t} ,
  \hspace{.2in} t > 0
\end{eqnarray}
where $S_0$ is the source strength and where we have put in
a factor of $t$
to make the time dependence continuous at $t=0$.
Using (12) this leads to
\begin{equation}
|f({\bf k})|^2 = \frac{C_0 k^2}
{(k^2+\alpha^2)^2(\omega_k^2+\gamma^2)^2 \omega_k^2}
\end{equation}
where $C_0$ is a strength and where we have multiplied
$f$ by $k$ to model the p-wave nature of pion emission.
With this form, the integrals for the mean number of pions
and for the mean energy (17) and (23) cannot both be done
analytically, but they can be easily evaluated numerically.
This requires a choice of the parameters $\alpha$, $\gamma$
and $C_0$.  The last can be fixed by requiring that the
average energy be the energy released in annihilation.
In units of the pion mass ($\mu =1$), and for annihilation at rest,
this energy is 13.87.  For the range parameters, we take
$\alpha =\gamma=2$.  This corresponds to an annihilation
region with a time and distance scale of half a pion Compton
wave length - a reasonable size.  It is not the purpose of this
paper to make a careful study of the source function and to fit
it to data.  Rather we simply want to demonstrate that a quite
reasonable choice for functional form and size yields a
correspondingly reasonable account of the data.
Thus we have not made an exhaustive parameter search.
With our choice for parameters we find that (17)
gives for the average pion number $\hat{N}=6$.  With the same
choice of parameters, the average pion number
calculated with four-momentum conservation imposed as in
(50) gives  $\hat{N}=6.4$.  This is certainly within $1/\hat{N}$
of the unconstrained value, as promised. Also $\hat{N}$ about 6
agrees with the data,
for reviews see \cite{S&S}, \cite{DoverGuts},
particularly when it is realized that
we are only modeling that part of annihilation that goes into uncorrelated
pions, and not the part that goes into other mesons that subsequently
decay into pions.  Since these typically give few pions, the
average number without them will be somewhat higher than
the  total average number.

Although the average number from the Poisson distribution and
the distribution constrained by four-momentum conservation come
out very close, the actual probability distributions in pion
number are rather different.  In Fig. 1 we show the probability
of finding $m$ pions as a function of $m$  for the Poisson and
for the ``m-sum" cases (as defined in (47) and (48)).
We see that the constrained case has a far narrower distribution
than the Poisson.  For the Poisson distribution
the $\sigma$ ($\sigma^2$ is
defined in (24)) is
$\sigma = \sqrt{\hat{N}} = 2.45$. For
the distribution constrained by four-momentum conservation we
find $\sigma=0.88$, which is quite close to the
value deduced \cite{Ches} from experiment,
$\sigma=1.02$. This value has a statistical error of
at least five percent and an additional unknown error
from modeling the distribution of neutrals.
In fact the entire pion multiplicity distribution from the
constrained calculation shown in Fig. 1 is quite close to
the experimental distribution \cite{Ches}, \cite{DoverGuts}.
Our constrained
distribution is also indistinguishable (on the scale of the
figure) from a Gaussian distribution with the same average
number and same variance. That is what a statistical model
would give.

One can also use the combination of iso-spin projection
and four momentum projection of Sect. 2.7 to  calculate
the pion number distributions.  They are shown in Fig. 2a and
2b for $I=0$ and $I=1$, $I_z=0$ respectively.  Recall
that for $I=0$ only an even number of pions is possible
while for $I=1$ there is only an odd number. The data is
an appropriate average over these two, and again strongly
resembles our figures \cite{Ches}, \cite{DoverGuts}.
Annihilation into vector mesons must be added to evade
this odd-even effect.

Let us now turn to a comparison of our picture with experiment
for the average number of pions by iso-spin type.
We compare with the ratios reported in Sedlak and Simak \cite{S&S}.
To do this we need to estimate the relative population of
$I=0$ and $I=1$ in proton antiproton annihilation at rest.
We do so following the work of Locher and Zou \cite{LZ}.
Then using the charge ratios in states of good iso-spin we
found in Sect. 2.5, and for vanishing total third component of
iso-spin,
we find $\hat{N}_0/\hat{N}_{+} =1.53$
with an error of about ten percent
to be compared with the experimental number reported as
$1.27 \pm .14$ \cite{Ches}, \cite{S&S}.
This is remarkable agreement.
Note that these ratios do not depend at all on the details of
our parameters, but do depend on the structure of the
coherent state picture.
In particular, it should be noted that the
excess of $\pi^0$'s found experimentally comes out
naturally in our treatment.
Given $|f({\bf k})|^2$, we can calculate charge
averages including four-momentum conservation as
in Sect. 2.7.  Using the same parameters,
we can find the average pion number by charge
type and the variance in the various iso-spin states.
These are shown in Table 1.  The $I=1$, $I_z=1$ case
shows the effect of charge conservation, for $I_z=1$,
the average number of $\pi^+$ is one larger than the
average number of $\pi^-$.
We also see that with four-momentum conservation the ratio of
$\pi^0$ to $\pi^+$ in the
$I=1$, $I_z=0$ case is somewhat larger than in the unrestricted
case.

To make our formalism a real theory of the
annihilation process we need to  model
annihilation
in space and
time as a classical source of pions.
We have presented a very simple example
as an orientation.  More sophisticated
models of the pion source could be made based on a
dynamical theory of the annihilation process, for
example the Skyrme model calculation \cite{Sommermann},
or on nucleon-antinucleon potential model calculations.
These could include the spatial asymmetry seen in these
annihilation calculations \cite{Sommermann}. This asymmetry
would lead to a different momentum distribution for
the pions.
Alternately one could use (22) to extract
the form of $f({\bf k})$
from the single pion momentum distribution data.
We must also model the substantial part of annihilation
that goes through meson resonances.
We plan to return to these ideas in later work.

Given a form for $f({\bf k})$, one can calculate not
just single pion averages, but also
correlations \cite{Boal}, \cite{Song} and joint momentum
distributions for multiple pions.  The great advantage
of the pion coherent state approach is that it makes
detailed, specific and rather restrictive predictions
for these observables and thus is easily verifiable,
or perhaps falsifiable.

\section{Future Work and Summary}

Clearly much work remains to be done to develop and test
the coherent state approach
developed here for pions radiated from nucleon
antinucleon annihilation. Predictions for
higher moments of pion
distributions from nucleon antinucleon states
of sharp iso-spin have to be confronted with data.
Correlations have to be studied.
Theories of nucleon antinucleon interaction and of the
development of the annihilation region have to be studied
and interpreted in terms of a source of the pion radiation.
It may be possible, for example, to model annihilation
classically in the Skyrme model including the
non-linear interactions of the pions, and  the
effects of the annihilation back on the source.
Such a description would lead, far enough away from the
annihilation region, to a free, non-interacting pion
wave.  This wave could then be the source of the
coherent state.  This approach is far simpler than a
full quantum theory of the annihilation.
Beyond this study of the annihilation process, we must
study the effect of other quantum numbers such as
G-parity and angular momentum on annihilation
in the coherent state description.
Finally a better understanding of the relation
between annihilation into pion radiation and into
other mesons is needed. We are exploring all these
questions. Should these detailed descriptions
fail, it would be interesting to go beyond the
coherent state picture by including squeezed states.

One can also imagine other applications
of the general methods advocated here.  The basic
idea is to look for any fast coherent processes
in which the total energy radiated is large
compared with the mass of the radiated quanta
and in which the interaction among those quanta
or by the quanta back on the source can be neglected.
Then the radiation process is described classically,
and the subsequent wave quantized using coherent states.
Conserved quantum numbers have to be imposed on these
states as we have imposed iso-spin.  One can then
explore averages, moments, momentum distributions
and their relation to the radiating source, all as we
did for annihilation. Greiner, Gong and Muller
\cite{Greiner} have discussed one such possible application in
heavy ion collisions, there are many more.  There may
be similar opportunities in the description of hadronization
in QCD jets \cite{Doks}.
Other examples will occur to the reader.
It may even be possible to lift the restriction of
no self interaction of the field or of the field back
on the source so long as one can treat these classically
and then use coherent states to quantize only in the radiation
zone. Many QCD processes, like annihilation,
may best be treated by starting in the classical non-perturbative
domain of QCD.  It may also be possible to extend these
ideas to heavy ion collisions in general \cite{Blaz},
where one could take into
account the statistical nature of the process by
introducing a density matrix and averaging over an
appropriate ensemble of coherent states.  Such
methods are well known in quantum optics \cite{glauber},
and may be an attractive alternative to the
thermodynamic approach to heavy ion collisions.

In summary we have seen that the
large energy released and
the possibility of very short reaction time suggest
that a classical, coherent pion wave pulse is radiated
in nucleon antinucleon annihilation.  This wave can
be quantized using the method of coherent states.
{}From this description averages and higher moments of
pion distributions in annihilation can be calculated.
In particular by imposing iso-spin
and four-momentum conservation, interesting
correlations among pion charge types are easily calculated.
Momentum distributions contain information on the spatial
and temporal distribution of the pion wave source. Preliminary
calculations show that all these features obtained
from the coherent state picture agree
with the data and therefore suggest
that the coherent classical pion pulse picture of annihilation
is valid for that part of annihilation
that goes directly into pions.
We also speculate on other applications
of classical meson waves and their subsequent quantization.

\section*{Acknowledgments}

RDA, FC, and JPD  would like to thank
the theory group of the Division of Nuclear
and Particle Physics at
the Paul Scherrer Institute
and for providing a very
pleasant environment in which much of this work was done.  RDA
and BS are
supported in part by the United States National Science Foundation.
The Division de Physique Th\'eorique is a Research Unit of the
Universities Paris 11 and 6 associated to CNRS.
RDA and BS  would like to thank Dr. R. Hollebeek for bringing
reference \cite{B&P} to our attention.

\newpage
\begin{flushleft}
Figure Captions \\
\vspace{1in}
Figure 1. The probability distribution, $P_m$, for finding $m$
pions from nucleon antinucleon annihilation at rest.  The open
circles are the Poisson distribution and the solid squares are
the distribution with the constraint of four-momentum
conservation imposed.  That distribution cannot be
distinguished (on the scale of the figure) from a
Gaussian of the same mean and variance.  \\
\vspace{.3in}
Figure 2. The probability distribution, $P_m$, for finding $m$
pions from nucleon antinucleon annihilation at rest with iso-spin
projection. In all cases the constraint of four-momentum
conservation has been imposed.
Figure 2a is for $I=0$ and Figure 2b for $I=1$, $I_z=0$. The
open circles are the probabilities with iso-spin projected and the solid
squares without iso-spin projection. The solid squares are the same
as the solid squares in Figure 1. Note that in our model, for $I=0$
only an even number of pions is possible and for $I=1$ only an odd number.
This makes the iso-spin projected distributions appear peaked.
\end{flushleft}
\newpage

\begin{flushleft}
Table 1: The mean pion number and variance, and the mean pion
number by charge type
from nucleon antinucleon annihilation
at rest in each of the three iso-spin states with the constraint of
four-momentum conservation included.
\end{flushleft}

\begin{table}
\begin{tabular}{|c|ccccc|}
Iso-spin state    &  $\hat{N}_{+}$ & $\hat{N}_{-}$ & $\hat{N}_{0}$
& $\hat{N}$ & $\sigma$ \\   \hline
$I=0, I_z=0$ &  2.07 & 2.07 & 2.07 & 6.21 & 0.82 \\  \hline
$I=1, I_z=0$ &  1.08 & 1.08 & 4.24 & 6.4  & 0.95 \\  \hline
$I=1, I_z=1$ &  3.15 & 2.15 & 1.08 & 6.38 & 0.95
\end{tabular}
\end{table}

\end{document}